\documentclass[prl, aps, 12pt, showkeys, showpacs,tightenlines] {revtex4}
\usepackage{amsmath}
\usepackage{epsfig}
\begin{document}
\title{Entropy and Entropy Production in Some Applications}
\author{Yong-Jun Zhang}
\email{yong.j.zhang@gmail.com}
\affiliation{Science College, Liaoning Technical University, Fuxin, Liaoning 123000, China}

\begin{abstract}
By using entropy and entropy production, we calculate the steady flux of some phenomena. The method we use is a competition method, $S_S/\tau+\sigma={\it maximum}$, where $S_S$ is system entropy, $\sigma$ is entropy production and $\tau$ is microscopic interaction time. System entropy is calculated from the equilibrium state by studying the flux fluctuations. The phenomena we study include ionic conduction, atomic diffusion, thermal conduction and viscosity of a dilute gas. 
\end{abstract}

\maketitle

\section{Introduction}
Transport phenomena, including electrical conduction, thermal conduction, viscosity and diffusion, are all about non-equilibrium states. A non-equilibrium state always tends to relax to the equilibrium state. But an opposite tendency can also exist if there is an external force, because an external force always induces its conjugate flux to tend to become greater. The two tendencies compete with each other and the compromise is a steady flux. However, their strengths seem measured in different quantities: one is system entropy, the other entropy production. In order to make them comparable, a time parameter $\tau$ must come in.

$\tau$ relates to microscopic interactions. For instance, it can be the mean molecular collision time, $\tau=\lambda/\bar{v}$. During such a $\tau$, a molecule averagely collides once. The collision result is non-deterministic. So randomness arises, which is then described by system entropy. A system entropy is thus associated with a $\tau$ as well as a series of discrete processes. By using $\tau$, system entropy and entropy production can be compared. There are three methods to do so, see Tab \ref{three_solutions}. 
\begin{table}[htbp]
\begin{tabular}{l|l|l|l}\hline
        & {\bf tendency for a flux to relax } & {\bf tendency for a flux to increase } &  \\ \hline
            & system entropy $S_S$         & entropy production $\sigma$ \\
             & $S_S-S_{S0}\propto -J^2$        & $\sigma\propto J$\\ \hline
method 1    & relaxation entropy production         & usual entropy production \\ 
             & $J(t)=Je^{-t/\tau}$        &  &\\ 
             & $\left.\frac{dS_S(t))}{dt}\right|_{t=0}$        & $\sigma$ &  $\left.\frac{dS_S(t))}{dt}\right|_{t=0}=\sigma$\\ \hline
method 2                       & system entropy                                                       & environment entropy \\
             & $S_S$        & $S_E=S_{E0}+\tau\sigma$ & $S_S+S_E =$ maximum \\ \hline
method 3                        & system entropy and $\tau$                                                        & entropy production \\
             & $S_S/\tau$           & $\sigma$ & $S_S/\tau+\sigma=$ maximum\\ \hline
\end{tabular}
\caption{Methods to calculate steady flux $J$.
}
\label{three_solutions}
\end{table}

The first method is an entropy production method \cite{yjzhang,yjzhang2,yjzhang5}. In this method, one constructs a new entropy production from the system entropy, by considering that a given flux relaxes at the maximum rate. The corresponding relaxation time is $\tau$ which now is interpreted as the shortest possible relaxation time. We can do this because the lower limit of the relaxation time should be the microscopic interaction time. Thus, a new entropy production is constructed, which is then compared with the other. When they are equal, the steady flux is found. Note that this method needs to be used together with two other principles: the steepest entropy ascent principle (SEA) \cite{Beretta5}, and the maximum entropy production principle (MEP) \cite{Paltridge2,Ziegler,MEPP,Dewar1}.

The second method is an entropy method \cite{yjzhang5}. In this method, one constructs a new entropy from the entropy production, by considering that the entropy producing process is a series of discrete processes. Each process lasts a time $\tau$. The newly constructed entropy is then added to the other. When the total entropy is maximum, the most probable flux is found, which is also the steady flux.

The third method is to be studied in this paper. It is a competition method, which leaves the system entropy and the entropy production in their original forms. In this method, it is the system entropy, rather than the entropy production, that is associated with discrete processes. A system entropy is always together with a $\tau$. They combine to form a single quantity, $S_S/\tau$, which truly measures the tendency of a flux relaxation.

We will present the competition method in a study of ionic conduction, and then extend it to other phenomena.

\section{ionic conduction}
\begin{figure}[htbp]
  \begin{center}
    \mbox{\epsfxsize=4.0cm\epsfysize=4.0cm\epsffile{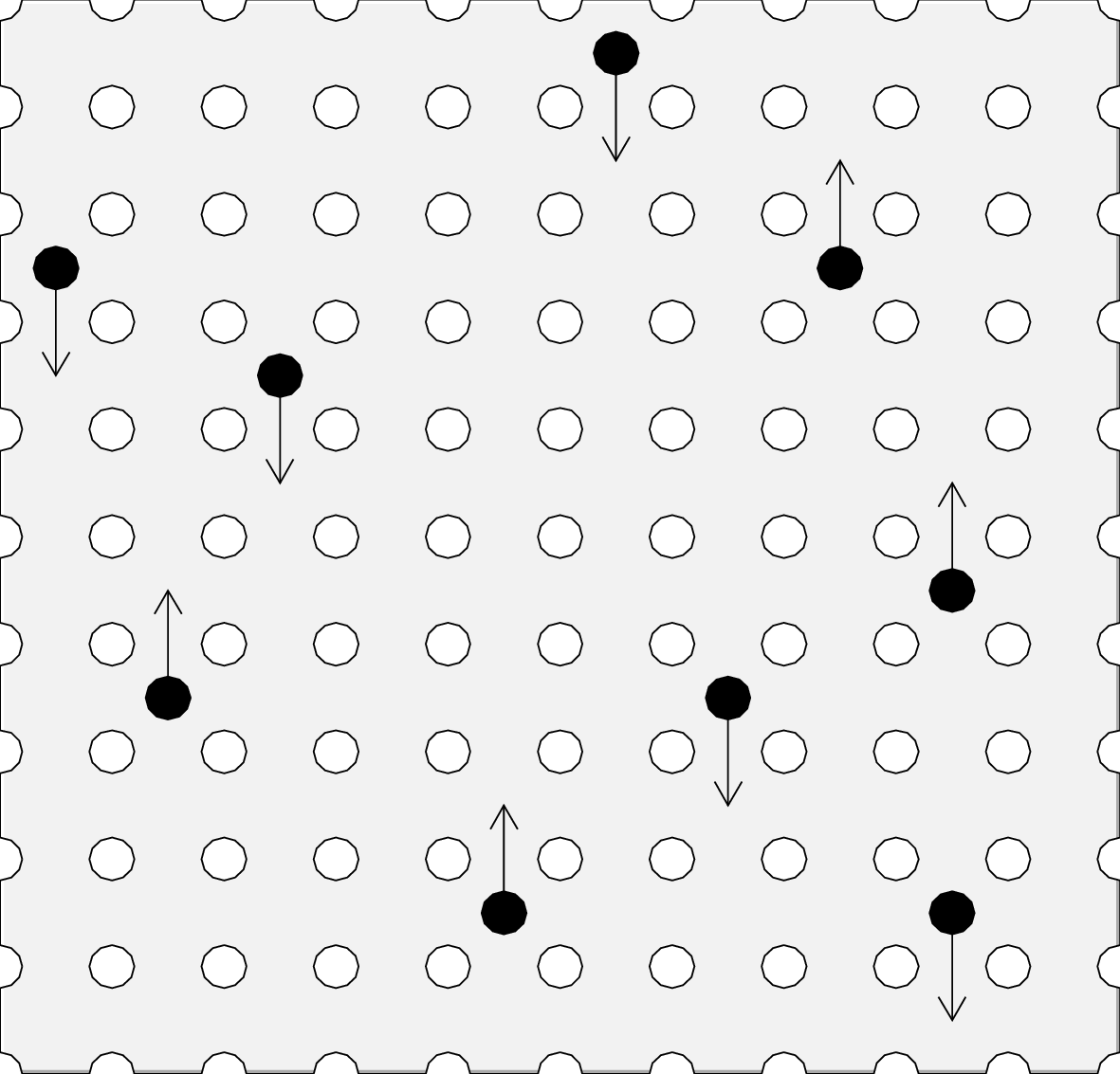}
        \ \ \ \ \ \epsfxsize=4.0cm\epsfysize=4.0cm\epsffile{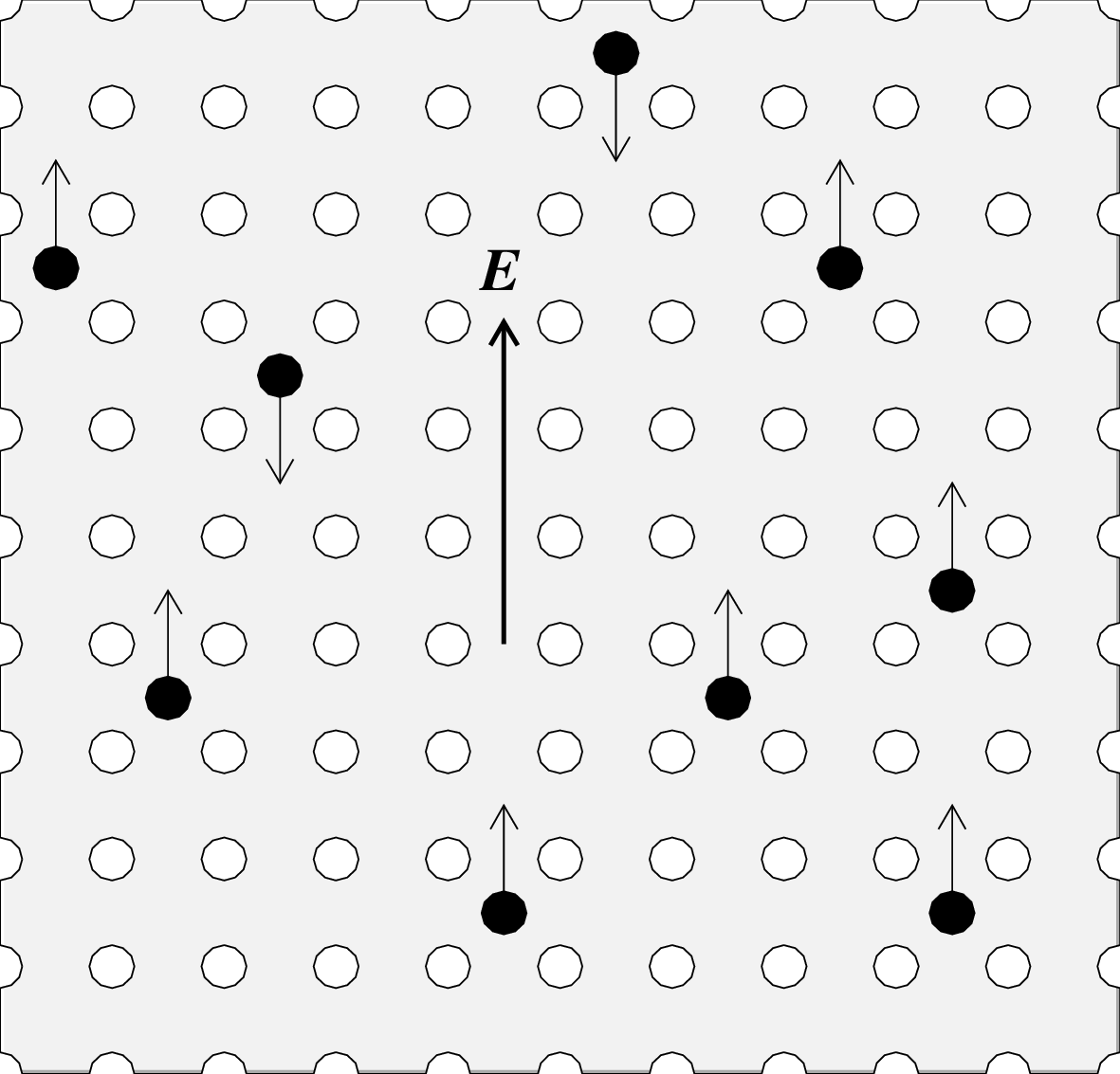}
        }
  \end{center}
\caption{
An ionic conductor. The white circles are a non-movable lattice, the black circles are interstitial ions. It has been idealized that each ion jumps once every $\tau$, either up or down. On the right, an external electric field $E$ exists and it has affected the ion jumps. We use $\{V,A,h,n\}$ to denote volume, cross section, height and ion number density.
\label{ions}}
\end{figure}

Ionic conduction is simple and ideal. An ionic conductor is shown in Fig. \ref{ions}.
The interstitial ions do not interact with each other in most of the time. This feature allows us to analyze one ion at a time and combine them later. Paper \cite{yjzhang5} has done this and we repeat the main points here. For one ion and one time interval $\tau$, there is
\begin{equation} \label{1ion}
\begin{array}{cl||c|c}\hline
         &    &\uparrow       &\downarrow \\ \hline
        \Omega_S& &1&1 \\
        J_V&=\sum\limits_{i=1}^Nqv_i&aq/\tau& -aq/\tau  \\
         Q&=J_VE\tau &aqE&-aqE  \\
        \Delta S_E&=Q/T&\frac{aqE}{T}&-\frac{aqE}{T}\\
        \Omega_E&\propto\exp\left(\frac{\Delta S_E}{k_B}\right) &\propto\exp(\frac{aqE}{k_BT})& \propto\exp(-\frac{aqE}{k_BT})\\
        P&\propto \Omega_E\Omega_S & \propto 1\times \exp(\frac{aqE}{k_BT})& \propto 1\times \exp(-\frac{aqE}{k_BT})\\\hline
\end{array}
\end{equation}
where
\begin{itemize}\setlength{\itemsep}{-3pt}
\item $\uparrow$ means an ion to jump up;
\item $\downarrow$ means an ion to jump down;
\item $\Omega_S$ is the number of microscopic system states for a given $J_V$;
\item $J_V$ is the overall flux of all ions, whose velocity is either $v=a/\tau \ {\rm or}\ -a/\tau$;
\item $N$ is the total number of ions;
\item $a$ is the lattice constant;
\item $q$ is the ion charge;
\item $Q$ is the heat generated during a $\tau$;
\item $\Delta S_E$ is the change of the environment entropy during a $\tau$;
\item $\Omega_E$ is the number of microscopic environment states;
\item $P$ is the probability for a given $J_V$ to appear during a $\tau$;
\item $\tau$ is the average time for an ion to make a jump, either up or down; the average time for an ion to make a jump up is $\frac{1}{\nu \exp\left(-\frac{\varepsilon}{k_BT}\right)}$, where $\nu$ is an effective vibration frequency and $\varepsilon$ is a potential barrier between two neighboring interstitial sites \cite{kinetic_diffusion}; for an ion to make a jump down, the discussion is the same; thus $\tau=\frac{1}{2\nu \exp\left(-\frac{\varepsilon}{k_BT}\right)}$.
\end{itemize}
For two ions and one time interval $\tau$, there is
\begin{equation} \label{2ions}
\begin{array}{c||c|c|c}\hline
        &\uparrow\uparrow       &\uparrow \downarrow {\rm \ or\ } \downarrow\uparrow&\downarrow\downarrow\\\hline
        \Omega_S&1&2&1\\
        J_V&2aq/\tau&0& -2aq/\tau\\
        Q &2aqE&0&-2aqE\\
        \Delta S_E&\frac{2aqE}{T}&0&-\frac{2aqE}{T}\\
        \Omega_E &\propto\exp(\frac{2aqE}{k_BT})&\propto1&\propto\exp(-\frac{2aqE}{k_BT})\\
        P&\propto1\times \exp(\frac{2aqE}{k_BT})&\propto2\times 1&\propto1\times \exp(-\frac{2aqE}{k_BT})\\\hline
\end{array} 
\end{equation}
For $N$ ions and one time interval, there is
\begin{equation} \label{Nions}
\begin{array}{c||ccccc}\hline
        \Omega_S &1&\cdots&C_N^k&\cdots&1\\
        J_V&N{aq}/{\tau}&\cdots&(2k-N)aq/\tau&\cdots& -Naq/\tau\\
        Q&NaqE&\cdots&(2k-N)aqE&\cdots&-NaqE\\
        \Delta S_E & \frac{NaqE}{T} &\cdots &\frac{(2k-N)aqE}{T}& \cdots & -\frac{NaqE}{T}\\
        \Omega_E & \propto \exp\left(\frac{NaqE}{k_BT}\right) &\cdots &\propto\exp\left(\frac{(2k-N)aqE}{k_BT}\right)& \cdots & \propto \exp\left(-\frac{NaqE}{k_BT}\right)\\
        P&\propto 1\times \exp\left({\frac{NaqE}{k_BT}}\right)&\cdots&\propto C_N^k\times \exp\left({\frac{(2k-N)aqE}{k_BT}}\right)&\cdots&\propto 1\times \exp\left(-\frac{NaqE}{k_BT}\right)\\\hline
\end{array} 
\end{equation}
where $k$ is the number of ions that jump up. The most probable $k$ can be calculated by one of the following equivalent formulas
\begin{equation} \label{statements}
        \begin{array}{rl}
                P 			&= {\rm maximum};\\
                k_B\ln P		&={\rm maximum};\\
                k_B\ln \Omega_S+\Delta S_E	&={\rm maximum};\\
                S_S+S_E				&={\rm maximum}.
        \end{array}
\end{equation}
Here $S_S=k_B\ln \Omega_S$ and $S_E=S_{E0}+\Delta S_E$. We call them system entropy and environment entropy. They are defined for one time interval $\tau$. Or we say that we focus on a single discrete process which lasts a time $\tau$. $S_{E0}$ is the initial environment entropy for a given process, which is a constant and can be dropped. So far, the second method in Tab. \ref{three_solutions} is recovered. 

\begin{figure}[htbp]
  \begin{center}
        \setlength{\unitlength}{1cm}  
        \centering      
        \begin{picture}(60,5)   
                \put(0,1){\epsfxsize=4.0cm\epsfysize=4.0cm\epsffile{ions_equilibrium_vz_E.eps}}
                \put(6,1){\epsfxsize=4.0cm\epsfysize=4.0cm\epsffile{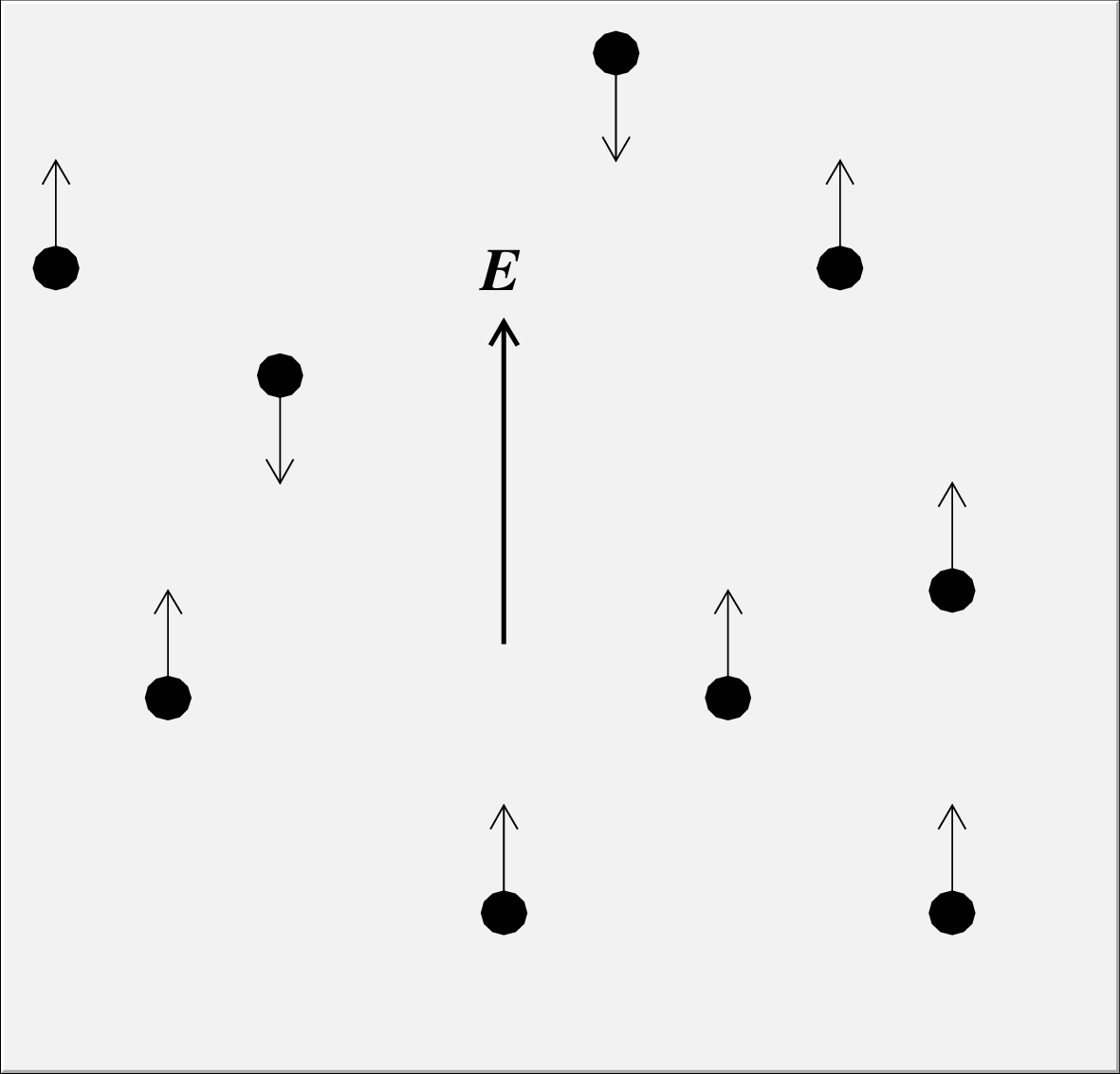}}
                \put(12,1){\epsfxsize=4.0cm\epsfysize=4.0cm\epsffile{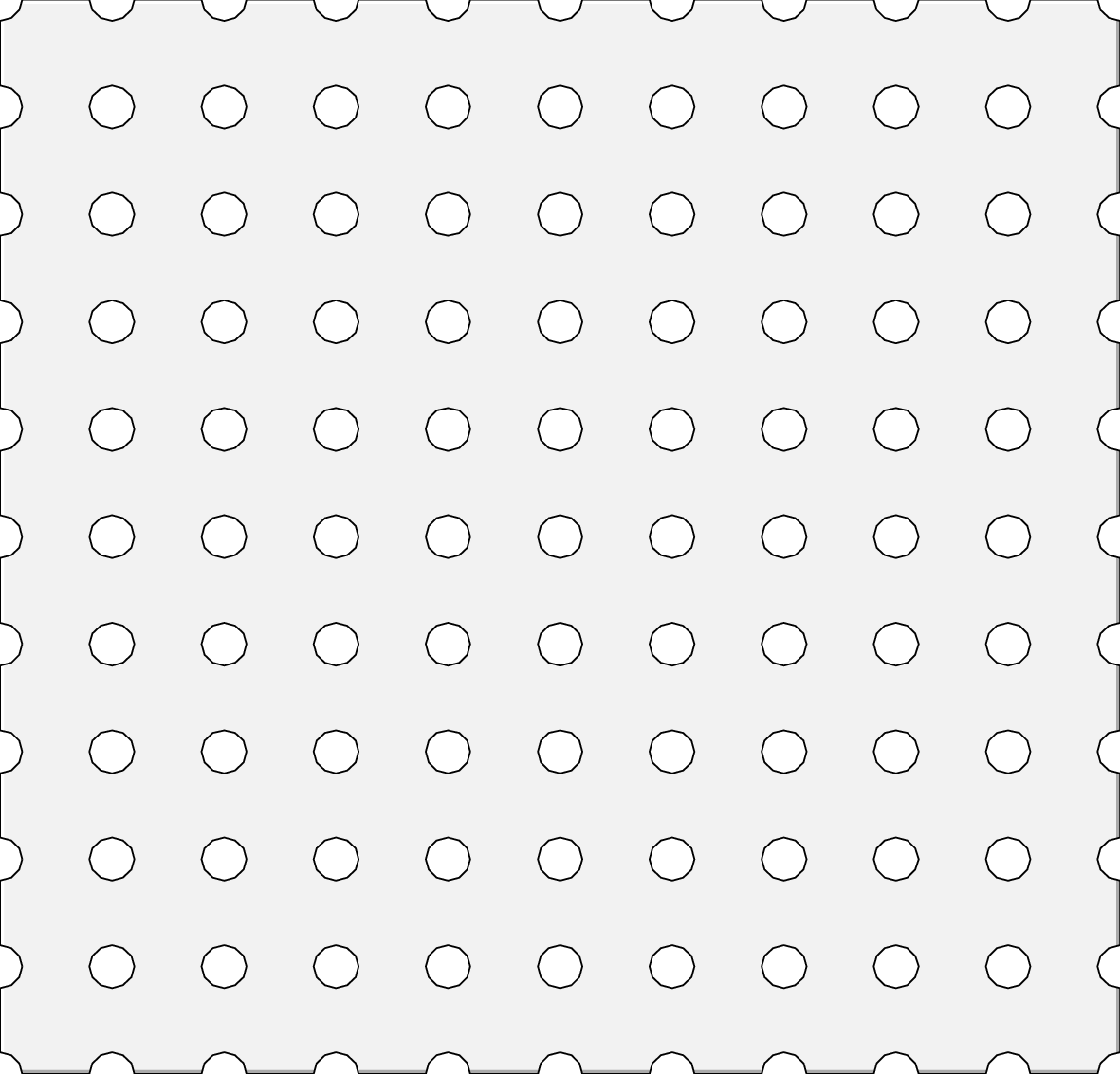}}
                \put(1,0) {conductor}
                \put(7,0) {the system}
                \put(12.0,0) {a part of the environment}
                \put(11,3){\vector(1,0){0.7}}
                \put(11,3){\vector(-1,0){0.7}}
                \put(10.8,3.3){Q}
        \end{picture}
  \end{center}
\caption{
A conductor being viewed as two parts. One part is the system, the other a part of the environment. The two parts exchange only heat $Q$.
\label{system}}
\end{figure}

Before going to the third method, we study a little of what the system is. In Fig. \ref{system}, a conductor is viewed as two parts: one is the system, the other a part of the environment. Correspondingly, the total entropy is divided into two parts too, as shown in Tab. \ref{table}. One is the system entropy which depends on only the flux. The other is a part of the environment entropy, which will increase, if there is an external force, at a rate called an entropy production.

\begin{table}[htbp]
\begin{tabular}{l|l|l}\hline
 	& {\bf system} & {\bf environment} \\ \hline
consists of		& ions	(and only about how& lattice;\\
			&	 they jump up and down)		& thermal vibrations of the ions;\\
			& 				& surroundings such as air \\ \hline
microscopic variables	&	$\tau$, $N$, $k$					& \\ \hline
macroscopic variables	&  $J_V=(2k-N)aq/\tau$						& $T$ 	\\ \hline
heat change in a $\tau$		 & & $Q=J_VE\tau$ \\ \hline
entropy change in a $\tau$		 & 	& $\Delta S_E={Q}/{T}$ \\ \hline
entropy production & 				& $\sigma={\Delta S_E}/{\tau}$ \\ \hline
number of microscopic states & $\Omega_S=C_N^k$				& $\Omega_E=\Omega_{E0} \exp({\Delta S_E}/{k_B})$ \\ \hline
entropy 		 & $S_S=k_B\ln \Omega_S$		& $S_E=S_{E0}+\Delta S_E$ \\ 
	&		& $\ \ \ \ \  =S_{E0}+\sigma \tau$ \\ \hline 
\end{tabular}
\caption{The system and the environment, and a process lasting a time $\tau$.
}
\label{table}
\end{table}

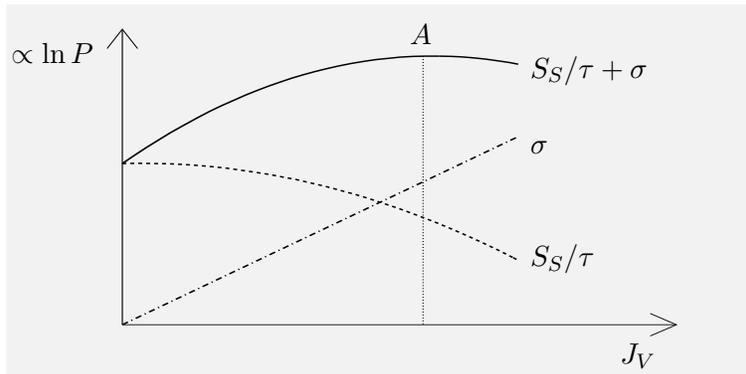
\begin{figure}[htbp]
  \begin{center}
        \setlength{\unitlength}{1cm}  
        \centering      
        \begin{picture}(60,5)   
                \put(0.5,0){\epsfxsize=10.0cm\epsfysize=5.0cm\epsffile{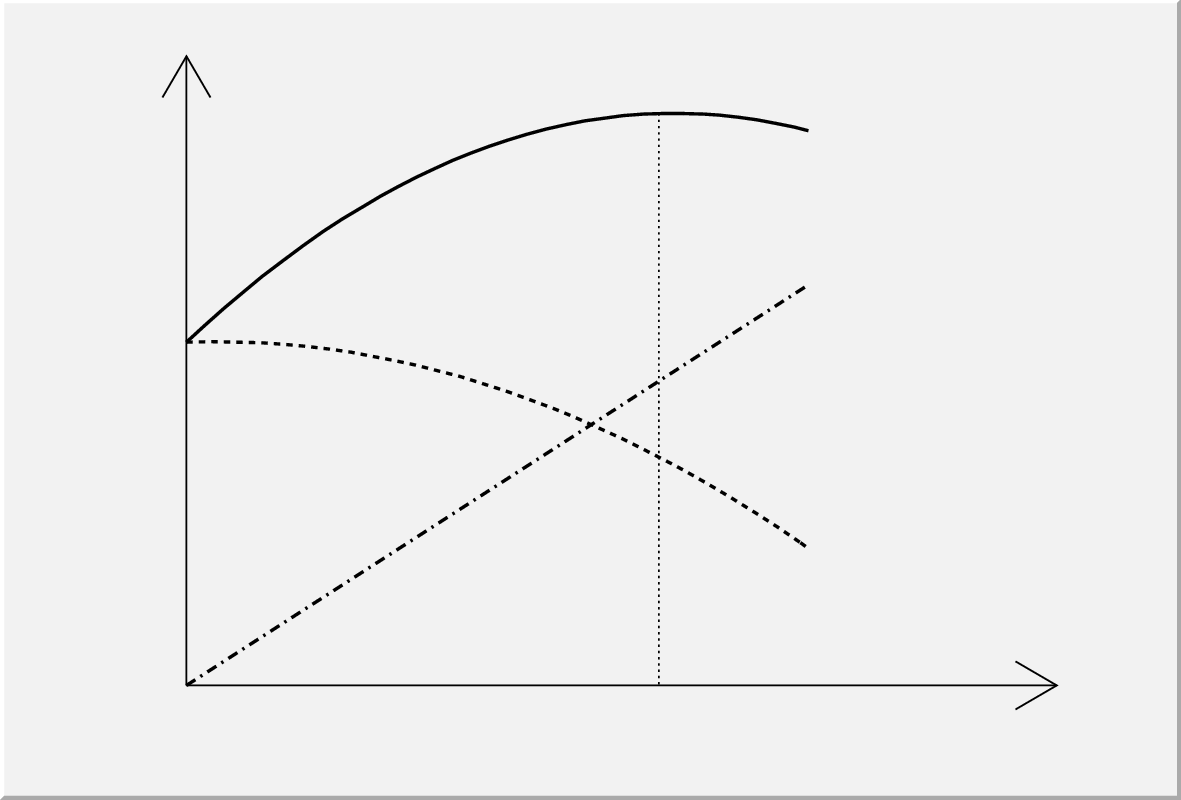}}
                \put(8.7,0.2) {$J_V$}
                \put(0.6,4.2) {$\propto\ln P$}
                \put(7.5,4) {${S_S}/{\tau}+\sigma$}
                \put(5.9,4.45) {$A$}
                \put(7.5,3) {$\sigma$}
                \put(7.5,1.5) {${S_S}/{\tau}$}
        \end{picture}
  \end{center}
\caption{
The competition method. At point A, ${S_S}/{\tau}+\sigma$ reaches the maximum, and the corresponding flux is the most probable flux, which is also the steady flux. $P$ is the probability for a $J_V$ to occur. The exact function is in the form like $\frac{k_B}{\tau}\ln P+y_0={S_S}/{\tau}+\sigma$ where $y_0$ is a constant.
\label{competition_method}}
\end{figure}

Tab. \ref{table} shows that $S_E=S_{E0}+\sigma\tau$. Plugging it into $S_S+S_E={\it maximum}$, we obtain $S_S+\sigma\tau={\it maximum}$. But we write
\begin{equation} \label{S_competition_method}
	\frac{S_S}{\tau}+\sigma={\rm maximum}.
\end{equation}
This is the third method that we call the competition method. See Fig. \ref{competition_method}. It has several features: It leaves all physical quantity in their original forms; It includes two terms, $S_S/\tau$ and $\sigma$, each represents a different tendency; It can be easily extended to include more terms to represent more fluxes or external forces.

Now let us introduce $s=S_S/V$ and $\sigma_s=\sigma/V$. Then we can write
\begin{equation}\label{ion_variation} 
	\frac{s}{\tau}+\sigma_s={\rm maximum}.
\end{equation}
This equation is particularly convenient for calculating the most probable $j$ which is also the steady $j$. Here $s$ is the system entropy per unit volume and $\sigma_s$ is the local entropy production whose notation is from \cite{sigma_s}. 

We next use the competition method to calculate the ionic conductivity. Tab. \ref{table} shows that 
\begin{equation} 
	\Omega_S=C_N^k.
\end{equation}
It is to be plugged into $S_S=k_B\ln \Omega_S$. We usually have $N\gg 1$ and $k\gg 1$. So we can use Stirling's approximation, $\ln n!\approx n\ln n-n$. And a variable $k$ relates to an electric flux $J_V$ in the form $J_V=(2k-N)aq/\tau$. So we get
\begin{equation}\label{S_S_ion} 
       	S_S=S_{S0}-\frac{1}{2}\frac{k_B\tau^2}{Na^2q^2}J_V^2.
\end{equation}
Using $s=S_S/V$, $J_V=jV$ and $N=nV$, we obtain
\begin{equation}\label{s_ion} 
       	s=s_0-\frac{1}{2}\frac{k_B\tau^2}{na^2q^2}j^2.
\end{equation}
As for the local entropy production, it is
\begin{equation} \label{sigma_s_electric}
	\sigma_s=\frac{E}{T}j.
\end{equation}
Plugging them into (\ref{ion_variation}), we obtain the most probable $j$,
\begin{equation} \label{j_ion}
        j=\frac{na^2q^2}{k_BT\tau}E.
\end{equation}
This is also the steady $j$, and it is the same as the result obtained in paper \cite{yjzhang5} which uses the other two methods.

We can further extract ionic conductivity, $\sigma_e=\frac{na^2q^2}{k_BT\tau}$. Thus we can rewrite (\ref{s_ion}) in the form $s=s_0-\frac{\tau}{2\sigma_eT}j^2$. This is the same as a result of work \cite{Jou2}, except that the parameter $\tau$ there was used as a relaxation time, just like how it is used in the first method in Tab. \ref{three_solutions}.

$J_V$ and $E$ can be extended to other kinds of fluxes and forces. We do this in the next three sections, following the steps listed in Tab. \ref{table2}. Some steps can be skipped, when the corresponding system entropy or the entropy production is known.
\begin{table}[htbp]
\begin{tabular}{l|l}\hline
{\rm\bf specify phenomenon} &{\rm identify what flux to study}\\
					&{\rm identify the conjugate force}\\ \hline
{\rm\bf calculate system entropy  $S_S$}     	&{\rm define system and set up the equilibrium state}\\
					&{\rm identify $\tau$ and idealize the system}\\
                        		&{\rm study  the  flux  fluctuations and calculate  the  system  entropy} \\ \hline
{\rm\bf calculate entropy production $\sigma$}   &{\rm identify the key part of the environment}\\
                        		&{\rm study its entropy properties} \\
                        		&{\rm for a given flux, calculate entropy production} \\ \hline
{\rm\bf use the competition method}   & {\rm calculate  the  most  probable  flux by $S_S/\tau+\sigma$= {\it maximum}} \\
                                & {\rm macroscopically, the  most  probable  flux is the  steady  flux }\\
                                & {\rm calculate transport coefficient}\\ \hline
\end{tabular}
\caption{The competition method and its steps
}
\label{table2}
\end{table}
%
%

\section{thermal conductivity of a dilute gas}
\begin{figure}[htbp]
  \begin{center}
        \setlength{\unitlength}{1cm}  
        \centering      
        \begin{picture}(60,5)   
                \put(0.5,0){\epsfxsize=5.0cm\epsfysize=5.0cm\epsffile{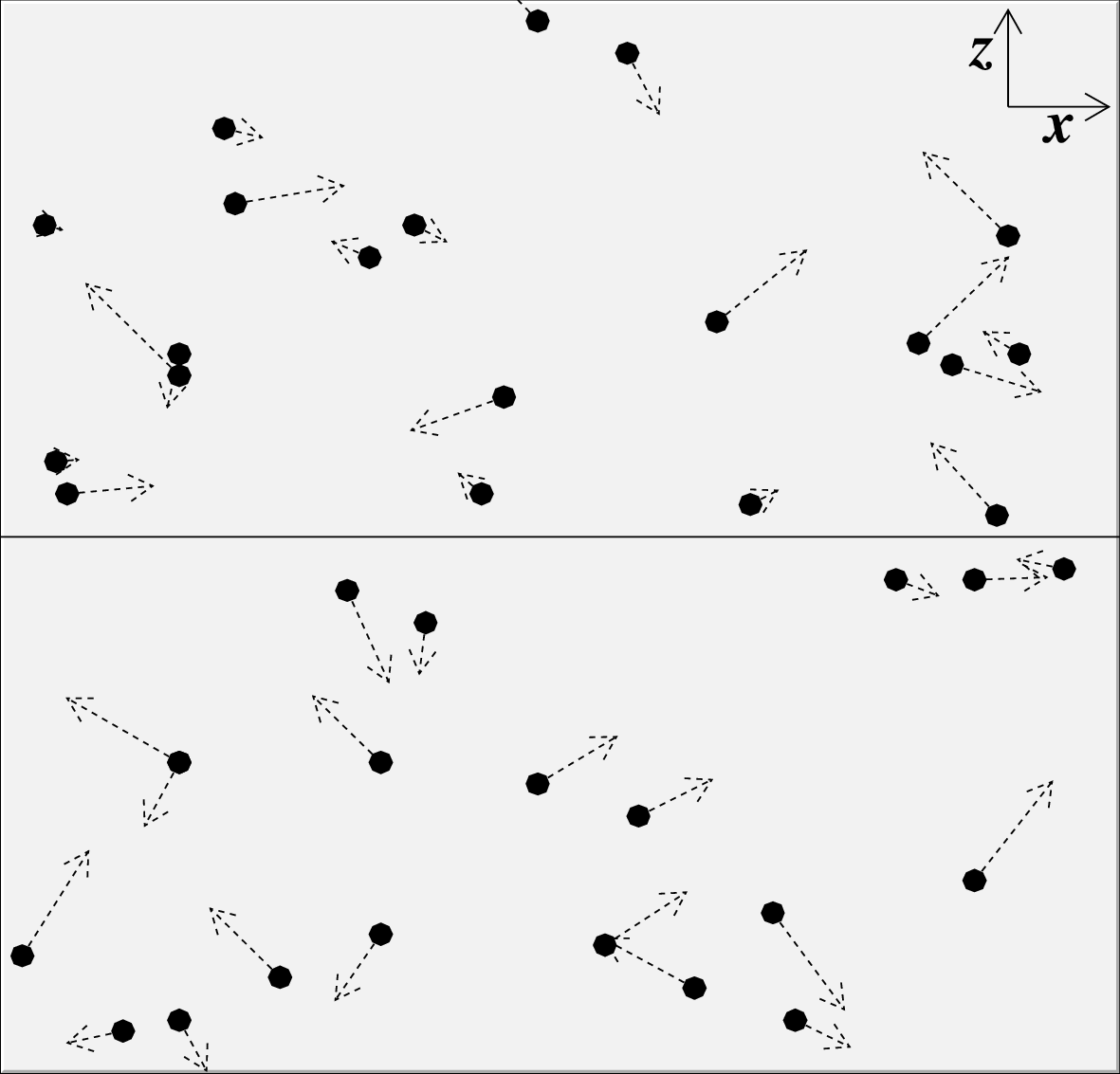}}
                \put(6.5,0){\epsfxsize=5.0cm\epsfysize=5.0cm\epsffile{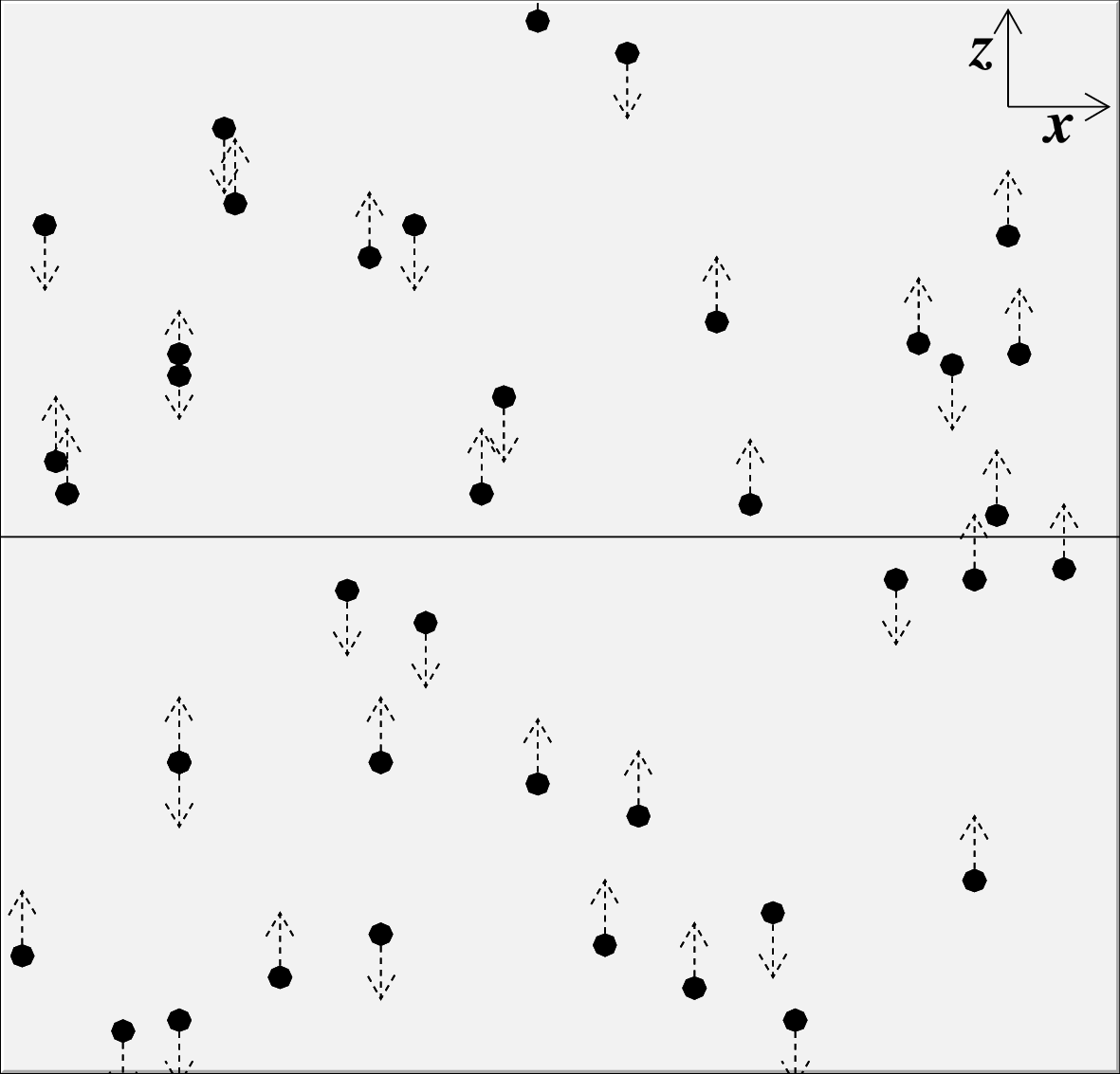}}
                \put(0,1) {$\lambda_z$}
                \put(0,3.5) {$\lambda_z$}
                \put(6,3.5) {$\frac{N}{2}$}
                \put(6,1.5) {$\frac{N}{2}$}
                \put(12,4.0) { $\uparrow k_1$ }
                \put(12,3.5) { $\downarrow N/2-k_1$ }
                \put(12,1.5) { $\uparrow k_2$ }
                \put(12,1.0) { $\downarrow N/2-k_2$ }
        \end{picture}
  \end{center}
\caption{ A system of a dilute gas to study thermal conductivity. It is in the equilibrium state and it has two layers. Each layer is of height $\lambda_z$ and contains $N/2$ molecules. On the right, the system is idealized that during each $\tau$ each molecule transfers the same energy $\varepsilon$, either up or down, to the next layer. 
\label{thermal}}
\end{figure}
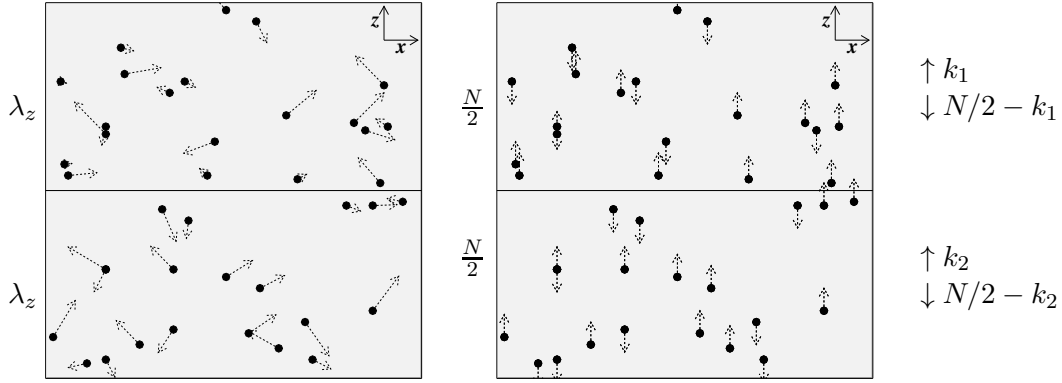

We next use the competition method to calculate the thermal conductivity of a dilute gas. The system entropy has been calculated in work \cite{yjzhang}. We summarize the main points below. The system is chosen as a dilute gas in the equilibrium state, as shown in Fig. \ref{thermal}. It includes two layers and each layer is of height $\lambda_z=\lambda/\sqrt{3}$. 

As for the time parameter $\tau$, it is
\begin{equation} 
	\tau=\frac{\lambda}{\bar{v}}.
\end{equation}
The system has been idealized that, during each $\tau$, each molecule transfers the same energy $\varepsilon$ either up or down. 
 The system consists of $N$ molecules, half of them in the up layer and half in the down layer. During a time interval $\tau$, there are $k$ molecules transferring energy up, the other $N-k$ molecules transfer energy down. The number $k$ fluctuates. A $k$ appears with the probability 
\begin{equation} 
	P(k)\propto C_N^k.	
\end{equation}
Each $k$ corresponds to a heat flux in the form
\begin{equation} 
	 J= k_2\varepsilon/\tau-(N/2-k_1)\varepsilon/\tau=(k-N/2)\varepsilon/\tau,
\end{equation}
where $k_1/k_2$ is the number of the molecules from the up/down layer transferring energy up. Then by using $S_S=k_B\ln \Omega_S$ and $\Omega_S\propto P(k)$, we obtain the system entropy
\begin{equation} \label{S_thermal} 
	S_S=S_{S0}-\frac{2k_B\tau^2}{N\varepsilon^2}J^2.
\end{equation}
Using $J=jA$, $N=2\lambda_zAn$ and $V=2\lambda_z A$, we obtain the system entropy per unit volume as
\begin{equation}\label{s_thermal} 
	s=s_0-\frac{1}{2}\frac{k_B\tau^2}{n\varepsilon^2\lambda_z^2}j^2
\end{equation}
where $n$ is the number density. We thus recover the system entropy obtained in work \cite{yjzhang}.

Note that the system entropy here is about the dynamical randomness arising from molecular collisions. Averagely, a molecule collides once every time interval $\tau$ every distance $\lambda_z$ in the direction of the temperature gradient. This is the reason why $\lambda_z$ is the length scale and $\tau$ is the time scale. 

We next consider entropy production. For thermal conduction, it is already known as
\begin{equation} \label{sigma_thermal}
	\sigma_s=-\frac{1}{T^2}\frac{dT}{dz}j.
\end{equation}
Think a dilute gas of height $\Delta z$ fixed in between two reservoirs with temperature $T$ and $T+\Delta T$ where $\Delta T=\frac{dT}{dz}\Delta z$. Then for a given heat flux $J$, the entropy production is $\sigma=J\left(\frac{1}{T+\Delta T}-\frac{1}{T}\right)$. In the limit $\Delta T\to 0$, and by using $J=jA$ and $\sigma_s=\sigma/V$ where $V=A\Delta z$, the above result is obtained.

Plugging (\ref{sigma_thermal}) and (\ref{s_thermal}) into (\ref{ion_variation}), we obtain the most probable $j$,
\begin{equation} 
	j=-\frac{n\varepsilon^2\lambda_z^2}{k_BT^2\tau}\frac{dT}{dz}.
\end{equation}
This is also the steady flux. Then the thermal conductivity is extracted as $\kappa=\frac{n\varepsilon^2\lambda_z^2}{k_BT^2\tau}$. It is the same as the kinetic result $\kappa=\frac{1}{3}nc\bar{v}\lambda$ with approximations $\varepsilon\sim cT\sim k_BT$, where $c$ is the molecular heat capacity. With the extracted $\kappa$, we can rewrite (\ref{s_thermal}) as $s=s_0-\frac{\tau}{2\kappa T^2}j^2$, to be the same as the result of works \cite{math, math2, Jou2}, except that the parameter $\tau$ was used there as a relaxation time. 

\section{Viscosity of a dilute gas}
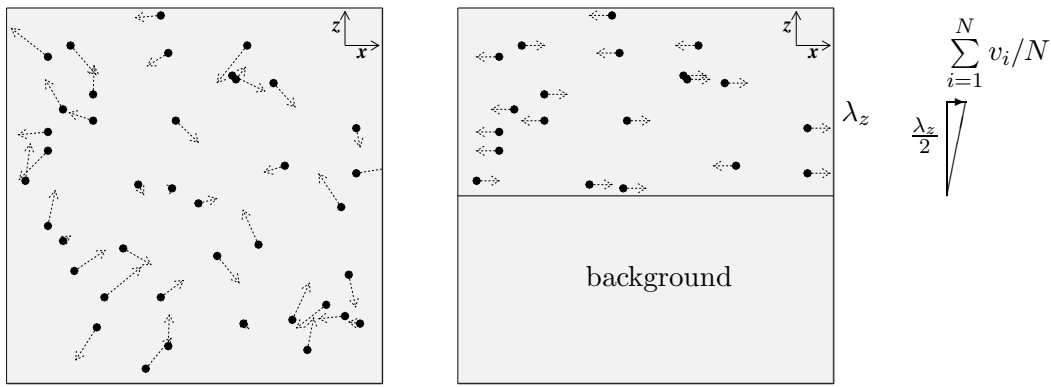
\begin{figure}[htbp]
  \begin{center}
        \setlength{\unitlength}{1cm}  
        \centering      
        \begin{picture}(60,5)   
                \put(0.5,0){\epsfxsize=5.0cm\epsfysize=5.0cm\epsffile{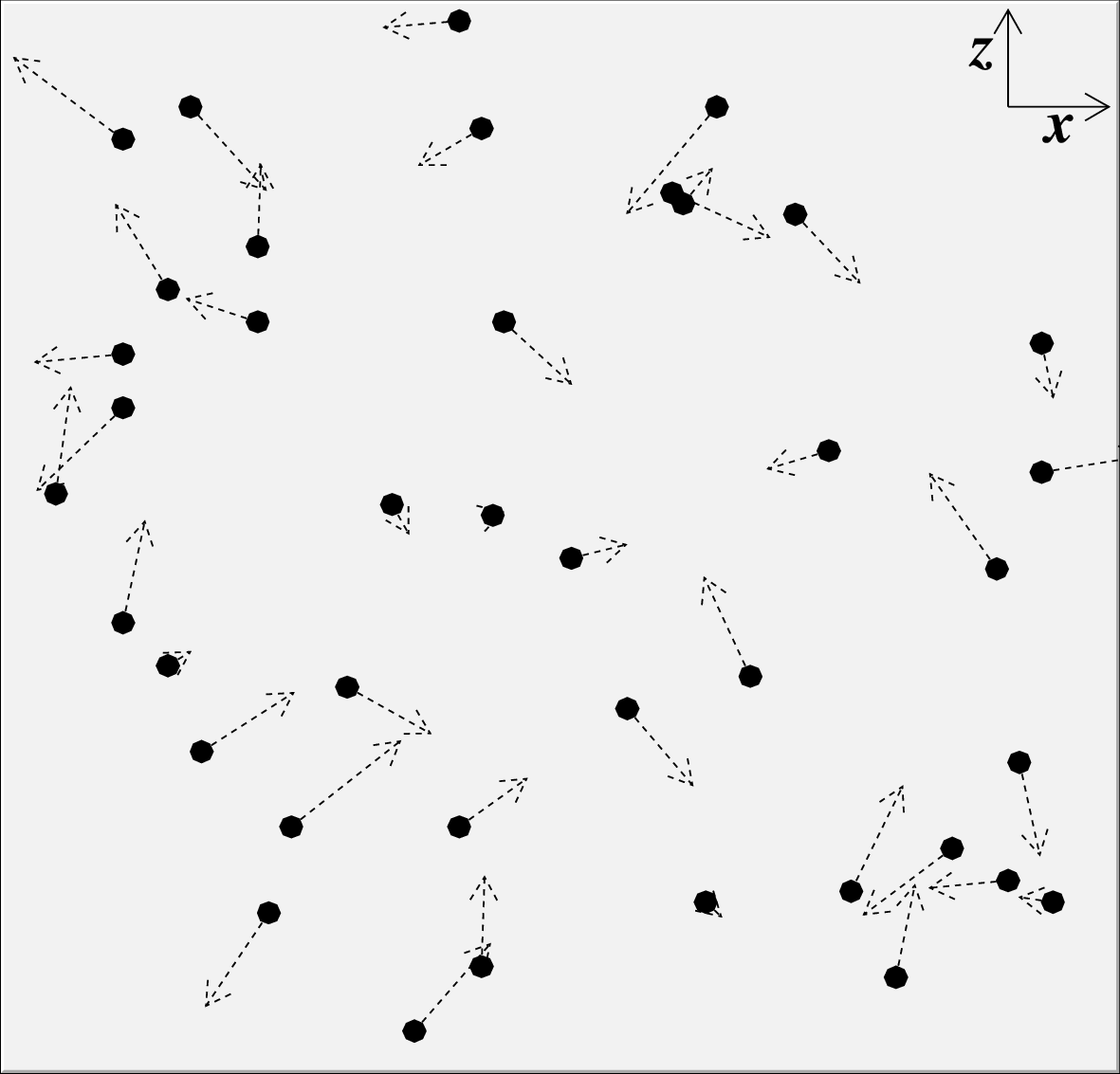}}
                \put(6.5,0){\epsfxsize=5.0cm\epsfysize=5.0cm\epsffile{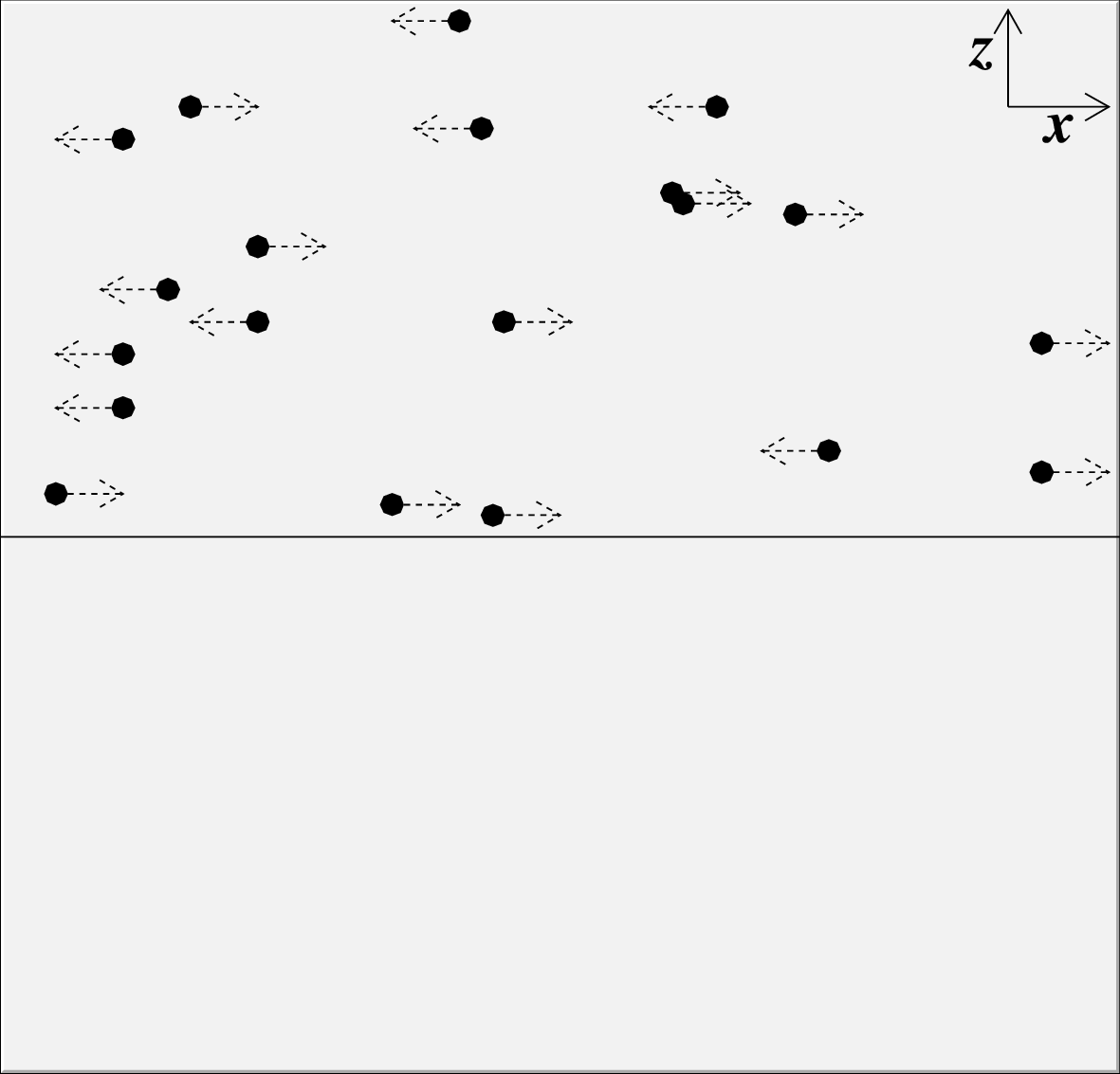}}
                \put(11.6,3.5) {$\lambda_z$}
		\put(13,2.5){\line(0,1){1.25}}
		\put(13,3.75){\vector(1,0){0.25}}
		\put(13,2.5){\line(1,5){0.25}}
		\put(13,4.3){$\sum\limits_{i=1}^N v_{i}/N$}
		\put(12.5,3.2){$\frac{\lambda_z}{2}$}
		\put(8.2,1.3){background}
        \end{picture}
  \end{center}
\caption{A dilute gas to study viscosity. The dilute gas is in the equilibrium state. We choose only a layer of height $\lambda_z$ as the system. The rest is the background. We study the relative velocity between the system and the background. The system is idealized that for a given $\tau$, each molecule moves either left or right with the same velocity $\bar{v}_z$ ($\bar{v}_x=\bar{v}_y=\bar{v}_z=\bar{v}/\sqrt{3}$).
\label{viscosity}}
\end{figure}

The system entropy of a dilute gas that carries a velocity gradient has been calculated in work \cite{yjzhang2}. We repeat here the main points. The system is chosen as a layer of dilute gas of height $\lambda_z$, as shown in Fig. (\ref{viscosity}). 

As for the time parameter $\tau$, it is
\begin{equation} 
 	\tau=\frac{\lambda}{\bar{v}}.
\end{equation}
The system has been idealized that, during a time interval $\tau$, each molecule moves either left or right with the same velocity $\bar{v}_z=\bar{v}/\sqrt{3}$.
 The total number of molecules is $N=nA\lambda_z$. During a time interval $\tau$, the number of those molecules that move right is denoted by $k$. The number $k$ fluctuates. A $k$ appears with the probability 
\begin{equation} 
	P(k)\propto C_N^k.
\end{equation}
A $k$ is associated with a velocity gradient in the form
\begin{equation} 
	\frac{dv_x}{dz}=\frac{\sum v_i/N}{\lambda_z/2}=\frac{k\bar{v}_z-(N-k)\bar{v}_z}{N\lambda_z/2}=\frac{(2k-N)\bar{v}_z}{N\lambda_z/2}.
\end{equation}
By using $k_B\ln P$, the corresponding entropy is obtained as
\begin{equation} 
	 S_{S0}-\frac{1}{8}\frac{k_BN\lambda_z^2}{\bar{v}_z^2}\left(\frac{dv_x}{dz}\right)^2.
\end{equation}
However, this is not the total system entropy, because so far only the overall velocity has been considered. For a uniform velocity gradient to appear, two more features need to be considered: (1) The system can be viewed as two layers and between them there exists a relative velocity; one of the two layers can be further viewed as another two layers... (2) The speed distribution has fluctuated away from the Maxwell speed distribution. 
 These two features have been considered in work \cite{yjzhang2}, and the system entropy has been approximately obtained there as
\begin{equation} 
	S_S\approx S_{S0}-\frac{1}{2}\frac{k_BN\lambda_z^2}{\bar{v}_z^2}\left(\frac{dv_x}{dz}\right)^2.
\end{equation}

In this paper, we start with this result. Using $s=S_S/V$, $V=\lambda_zA$ and $N=n\lambda_zA$, we obtain the system entropy per unit volume 
\begin{equation}\label{s_viscosity} 
	s= s_0-\frac{1}{2}\frac{k_Bn\lambda_z^2}{\bar{v}_z^2}\left(\frac{dv_x}{dz}\right)^2.
\end{equation}
This system entropy is about the configurational randomness arising from the molecular collisions. Averagely, a molecule collides once every time interval $\lambda/\bar{v}$ every distance $\lambda_z$ in the direction of the velocity gradient. So, the velocity gradient on length scale $\lambda_z$ fluctuates every time interval $\lambda/\bar{v}$. This is the reason why $\lambda_z$ is the length scale and $\tau=\lambda/\bar{v}$.

For viscosity, the local entropy production is known as
\begin{equation}\label{sigma_viscosity} 
	\sigma_s=\frac{\tau_{zx}}{T}\frac{dv_x}{dz}
\end{equation}
where $\tau_{zx}$ is the shear stress and $T$ is the temperature. To understand this, let us think of a dilute gas which is of temperature $T$, height $\Delta z$, cross section $A$, and fixed in between two plates: a stationary bottom plate and a top plate moving at velocity $v_x$ caused by a shear stress $\tau_{zx}$. Such a dilute gas, in each unit time, generates heat $\Delta Q=\tau_{zx}v_xA$, which is then absorbed by the dilute gas, though maybe further transferred somewhere else. The corresponding entropy production is $\sigma=\Delta Q/T$. By using $v_x=\frac{dv_x}{dz}\Delta z$ and $\sigma_s=\sigma/V$ where $V=A\Delta z$, the above local entropy production is obtained.

Plugging (\ref{sigma_viscosity}) and (\ref{s_viscosity}) into (\ref{ion_variation}), we obtain the most probable velocity gradient
\begin{equation} 
	\frac{dv_x}{dz}=\frac{1}{k_BTn\tau}\tau_{zx}.
\end{equation}
This is also the steady velocity gradient. So the viscosity is extracted as $\eta=k_BTn\tau=\frac{\pi}{8}\rho \bar{v}\lambda$, which is close to the kinetic result $\eta=\frac{1}{3}\rho\bar{v}\lambda$, where $\rho$ is the mass density. And we can rewrite the system entropy density (\ref{s_viscosity}) in the form $s=s_0-\frac{\tau \eta}{2T}\left(\frac{dv_x}{dz}\right)^2$, where the parameter $\tau$ can be interpreted as a relaxation time.

\section{atomic diffusion}
In this section we apply the competition method to study atomic diffusion. We reuse Fig. \ref{ions}, but set $E=0$, or replace all interstitial ions by neutral impurity atoms. The system entropy is the same as (\ref{S_S_ion}). But now we are not with the electric flux $(2k-N)aq/\tau$. Instead, we are with a diffusion flux
\begin{equation} 
	J_V=(2k-N)a/\tau,
\end{equation}
by which the system entropy is in the form,
\begin{equation} 
       	S_S=S_{S0}-\frac{1}{2}\frac{k_B\tau^2}{Na^2}J_V^2.
\end{equation}
The system entropy per unit volume is then
\begin{equation}\label{s_diffusion} 
       	s=s_0-\frac{1}{2}\frac{k_B\tau^2}{na^2}j^2.
\end{equation}

\begin{figure}[htbp]
  \begin{center}
    \mbox{\epsfxsize=10.0cm\epsfysize=1.33cm\epsffile{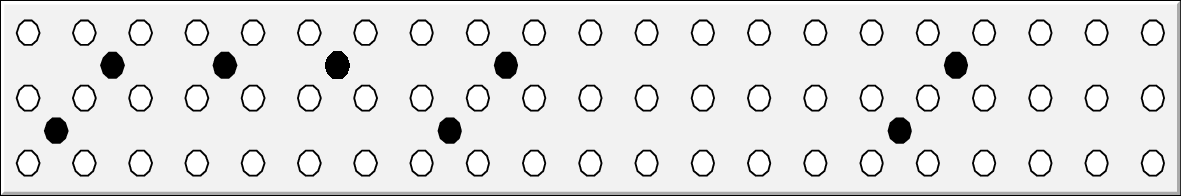}
        }
  \end{center}
\caption{
Two layers of impurity atoms and the background lattice. The lattice constant is $a$. The cross section is $A$, the volume is $V=2aA$. The number of atoms is $N$. The diffusion flux is $J$, which flows between the two layers. 
\label{atoms}}
\end{figure}

Concerning entropy production, we calculate it following the corresponding steps in Tab. \ref{table2}. We choose to study two layers of impurity atoms as shown in Fig. \ref{atoms}.
There are $N$ atoms and each atom has the same probability to be in the up layer or the down layer. When the up layer has $k$ atoms, the down layer has $N-k$ atoms. The number of the corresponding microscopic states is
\begin{equation} 
 	\Omega_E=C_N^k.
\end{equation}
$\Omega_E$ relates to atomic concentration gradient, $\frac{dn}{dz}$. 
 As long as we treat $\frac{dn}{dz}$ as an external force, $\Omega_E$ is a quantity of the environment. Thus, the environment entropy, or a part of it, is
\begin{equation} 
	S_E=k_B\ln \Omega_E=S_{E0}-\frac{2k_B}{N}\left(k-\frac{N}{2}\right)^2.
\end{equation}
When a diffusion flux $J$ flows between the two layers, the number $k$ changes. The environment entropy changes too, and the changing rate is entropy production. So the entropy production is
\begin{equation} 
	\sigma=\frac{dS_E}{dt}=-\frac{4k_B}{N}\left(k-\frac{N}{2}\right)\frac{dk}{dt}=-\frac{4k_B}{N}\left(k-\frac{N}{2}\right)J.
\end{equation}
Since $\frac{dn}{dz}$ is given, we write 
\begin{equation} 
	k-\frac{N}{2}=\frac{dn}{dz}\frac{a}{2}\frac{V}{2}.
\end{equation}
And by $N=nV$, we write  
\begin{equation} \label{sigma_diffusion_J}
	\sigma=-\frac{k_Ba}{n}\frac{dn}{dz}J.
\end{equation}
Note that $V=2aA$ where $A$ is the cross section. Then by using $\sigma_s=\sigma/V$ and $J=jA$, we obtain the entropy production per unit volume as
\begin{equation} \label{sigma_diffusion}
	\sigma_s=-\frac{k_B}{2n}\frac{dn}{dz}j.
\end{equation}

Plugging (\ref{sigma_diffusion}) and (\ref{s_diffusion}) into (\ref{ion_variation}), we obtain the most probable $j$,
\begin{equation} \label{j_diffusion}
	j=-\frac{a^2}{2\tau}\frac{dn}{dz}.
\end{equation}
This is also the steady flux. So the diffusion coefficient is extracted, $D=\frac{a^2}{2\tau}$, which is the same as the kinetic result \cite{kinetic_diffusion}. If we plug (\ref{j_diffusion}) into (\ref{sigma_diffusion}) and set $k_B=1$, we can write $\sigma_s\tau = \frac{D\tau}{2n}\left(\frac{dn}{dz}\right)^2$. This is the same as the result of work \cite{diffusion_entropy_production}, except that we have here an extra factor $\frac{1}{2}$ and the parameter $\tau$ there was interpreted as a relaxation time. Also, we can write the system entropy density in the form $s=s_0-\frac{k_B\tau}{4nD}j^2$, where the parameter $\tau$ can also be interpreted as a relaxation time.

The same result can also be obtained directly by formula (\ref{S_competition_method}). Compare Fig. \ref{atoms} with Fig. \ref{thermal}. They are the same if each layer contains the same number of particles. The system entropies are the same too, except that their variables are different, $\lambda_z\to a$ and $\varepsilon\to 1$. Since the system entropy for thermal conduction is (\ref{S_thermal}), the system entropy for atomic diffusion is $S_S=S_{S0}-\frac{2k_B\tau^2}{N}J^2$. Plugging it and (\ref{sigma_diffusion_J}) into (\ref{S_competition_method}), we obtain again (\ref{j_diffusion}).

\section{discussion}
\begin{figure}[htbp]
  \begin{center}
        \setlength{\unitlength}{1cm}  
        \centering      
        \begin{picture}(60,5)   
    		\put(0.5,0){\epsfxsize=9.0cm\epsfysize=5.0cm\epsffile{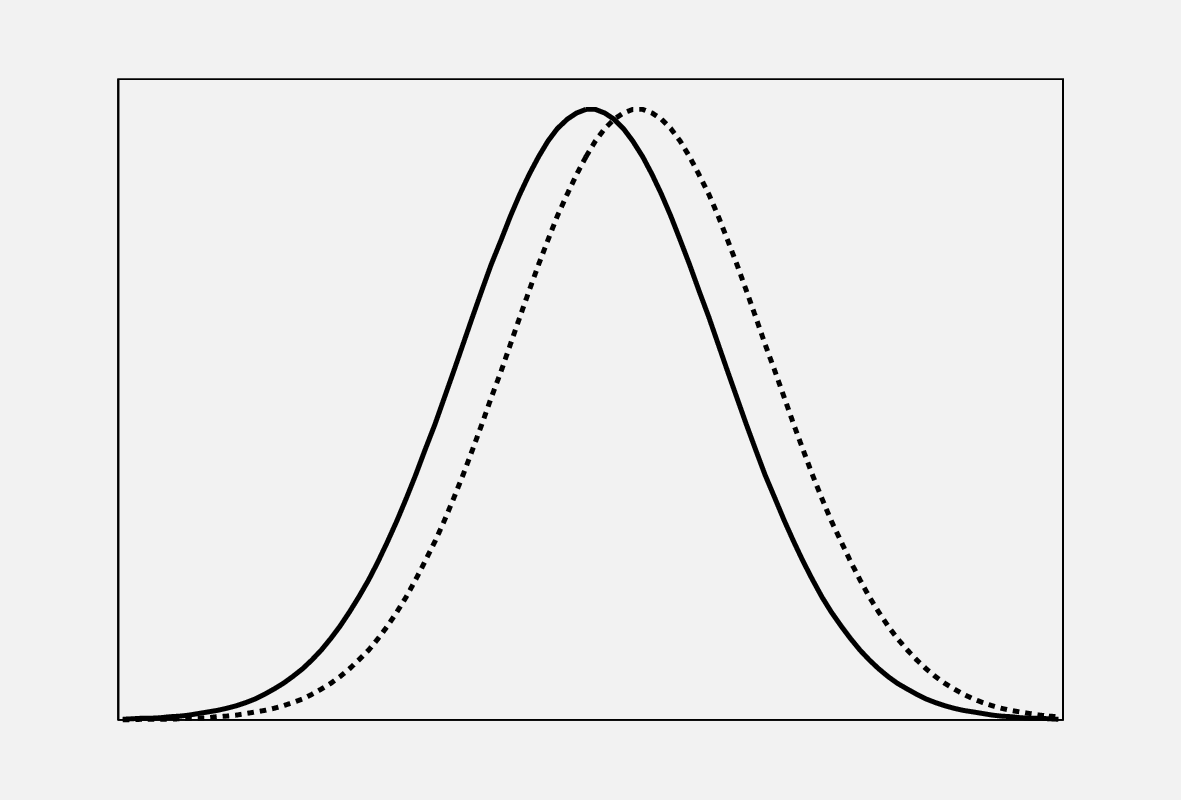}}
                \put(5,0.5){\line(0,1){4.}}
                \put(8.5,0.1) {$J_V$}
                \put(5,0.1) {$0$}
                \put(0.8,3.9) {$P$}
                \put(7.3,3.9) {$E=0$}
                \put(7.3,3.4) {$E\neq0$}
                \put(6.5,4){\line(1,0){0.5}}
                \put(6.5,3.5){$\cdots$}
                \put(6.575,3.5){$\cdots$}
        \end{picture}
  \end{center}
\caption{The fluctuations of a flux. The solid line is of the equilibrium state ($E=0$). The dashed line is of a non-equilibrium state ($E\neq 0$). The flux $J_V$ is defined as $\int_0^{\tau}J_V(t)dt/\tau$. A $J_V$ appears with probability $P$. The external force is $E$, but it can be extended to other kinds of force.
\label{fluctuation}}
\end{figure}
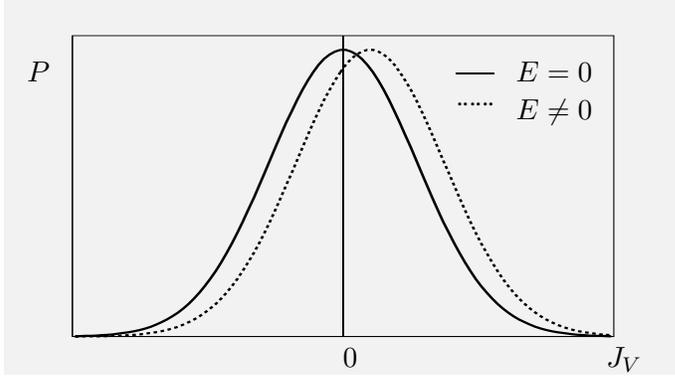

The system entropy is introduced to be against the environment entropy, or even directly against the entropy production. The system entropy can be calculated in two ways. One is to use $S_S(k)=k_B\ln\Omega_S(k)$ where $\Omega_S(k)$ is the number of microscopic states. The other is to use $S_S(J_V)= k_B\ln P(J_V)$ where $P(J_V)$ is the flux fluctuations shown in Fig. \ref{fluctuation} for the solid line.

For $S_S(J_V)$,  $J_V$ takes only discrete values that relate to $k$ in the form $J_V=(2k-N)aq/\tau$. But for $P(J_V)$, $J_V$ is usually treated as a continuous variable in order to have $P(J_V)=\frac{dP}{dJ_V}$ and $\int P(J_V)dJ_V=1$. Thus if one literally links $S_S(J_V)$ to $P(J_V)$, he will get some extra factors. Fortunately, all these factors will merge with $S_{S0}$ which is then dropped, or will not survive the variational procedure anyway. 

Concerning the dashed line in Fig. \ref{fluctuation}, it is the flux fluctuations for a non-equilibrium state. It has the same shape as the solid line, at least when $E\to 0$. This is the reason why they are associated with the same system entropy. Note that the dashed line is different from the solid line by a shift. This shift is exactly the shift of the most probable $J_V$ from zero.

In Tab. \ref{discussion}, we further compare the equilibrium state with a non-equilibrium state. Again, we see that they have the same system entropy in the limit $E\to 0$. We also see that the following formulas are equivalent,
\begin{equation}\label{s} 
        \begin{array}{rll}
                P(J_V,E)			&={\rm maximum},	&({\it statistical\ analysis})\\
		S(J_V,E)			&={\rm maximum},&({\it maximum\ entropy\ principle})\\
          	S_S(J_V)/\tau+\sigma(J_V,E) 	&={\rm maximum}. &({\it competition\ method})
        \end{array}
\end{equation}
The constant $S_{E0}$ has been dropped from the last formula.
\begin{table}[htbp]
\begin{tabular}{l|l|l}\hline
            & equilibrium  state       & non-equilibrium  state \\ 
                &$(E=0)$          & $(E\neq 0)$ \\ \hline
        flux fluctuations &  $P(J_V)$&  $P(J_V,E)$\\\hline
         entropy ($\propto k_B\ln P$) &  $S(J_V)$ &  $S(J_V,E)$ \\\hline
         entropy in the limit $E\to 0$ &  &  $S(J_V)+E\left.\frac{dS(J_V,E)}{dE}\right|_{E=0}$, or \\
        &  &  $S(J_V)+\tau\sigma(J_V,E)$ \\\hline
 introducing system entropy       & $S_S(J_V)+S_{E0}$  &  $S_S(J_V)+S_{E0}+\tau\sigma(J_V,E)$ \\\hline
 introducing environment entropy       &  &  $S_S(J_V)+S_E(J_V,E)$ \\\hline
\end{tabular}
\caption{The equilibrium state and a non-equilibrium state. The quantities are of a discrete process lasting a time $\tau$. The external force is $E$, which can be extended to other kinds of force.}
\label{discussion}
\end{table}

Now let us discuss the definition of $s$ in (\ref{ion_variation}). It has been defined as system entropy per unit volume. It can also be defined as a system entropy of a system that is of a unit volume. The two definitions are different, though their difference is small. To see this, let us study two systems:
\begin{equation}\nonumber
\begin{array}{l|l|l}\hline
            		& {\rm system\ of\ a\ unit\ volume}        & {\rm system\ of\ volume\ }V \\ \hline
        {\rm number\ of\ ions } 	&n          & N=nV \\ 
        {\rm   flux	}	& j 	&  J_V=jV \\
        {\rm number \ of\ ions\ jumping \ up}	& k 	&  kV \\
        {\rm number \ of\  microscopic\ states}	& C_n^k 	&  C_{nV}^{kV} \\
        {\rm  system \ entropy}	& \hat{s}=k_B\ln C_n^k 	&  S_S=k_B\ln C_{nV}^{kV} \\\hline
\end{array}
\end{equation}
One is of a unit volume, the other is of volume $V$. They share the same $j$. Note that $C_{nV}^{kV}\neq (C_n^k)^V$. So $S_S\neq V\hat{s}$. But $S_S-V\hat{s}$ is small. To see this, we need to use Stirling's series $n!\approx \left(\frac{n}{e}\right)^n\sqrt{2\pi n}$ instead of Stirling approximation $\ln n! \approx n\ln n-n$, which is not accurate enough in this case. Then we have
\begin{equation} 
	S_S-V\hat{s}=k_B\ln C_{nV}^{kV}-Vk_B\ln C_n^k=k_B\left[\frac{V-1}{2}\ln\frac{2\pi k(n-k)}{n}-\frac{1}{2}\ln V\right].
\end{equation}
We usually have $k\sim n/2$ and $n\gg 1$. So when $V>1$, we will have $S_S > V\hat{s}$. This reflects the fact that, a $V$-volume-system can not be viewed as $V$ copies of a unit-volume-systems, because the local flux fluctuates and is not necessarily all equal to $j$, even under the constraint $J_V=jV$. The quantity $S_S-V\hat{s}$ is in a scale of $\ln k$. $S_S$ is in a scale of $k\ln k$.  $S_S-V\hat{s}$ is relatively small. We can not even see it from (\ref{S_S_ion}), which is an approximation and does not include the corresponding terms. But the relation (\ref{ion_variation}) is exact, for which $s$ can be $\hat{s}$. To see this, let us start with (\ref{S_competition_method}) and study a system of a unit volume, $V\to 1$. Then we can write $J_V\to j$, $S_S\to s$, $\sigma\to \sigma_s$ and $N\to n$. After these replacements, (\ref{S_competition_method}) becomes (\ref{ion_variation}). (\ref{s_ion}) is obtained from (\ref{S_S_ion}) too.

Note that all system entropies look similar, including (\ref{s_ion}), (\ref{s_thermal}), (\ref{s_viscosity}) and (\ref{s_diffusion}). Actually we can introduce for each of them a dimensionless variable $\hat{j}$ and write them all in the same form 
\begin{equation} 
	s=s_0-\frac{1}{2}k_Bn\hat{j}^2, \ \ \ \ 
	S_S=S_{S0}-\frac{1}{2}k_BN\hat{j}^2.
\end{equation}
For ionic conduction, the dimensionless variable is $\hat{j}=\frac{j}{nqa/\tau}$ or $\frac{J_V}{Nqa/\tau}$. For thermal conduction, it is $\hat{j}=\frac{j}{n\varepsilon \lambda_z/\tau}$ or $\frac{J_V}{N\varepsilon \lambda_z/\tau}$. For diffusion, it is $\hat{j}=\frac{j}{na/\tau}$ or $\frac{J_V}{Na/\tau}$. For viscosity, it is $\hat{j}=\frac{dv_x}{dz}/\frac{\bar{v}_z}{\lambda_z}$ or $\frac{d\hat{v}_x}{d\hat{z}}$ with $\hat{v}_x=\frac{v_x}{\bar{v}_z}$ and $\hat{z}=\frac{z}{\lambda_z}$. Note that $N$ means $N$ particles and one time interval $\tau$. But the result will be the same if $N$ is to mean one particle and $N$ time intervals, which is a case of a stochastic process. This idea can go further: With $N=N_1N_2$, $N$ can mean $N_1$ particles and $N_2$ time-intervals, which is a case that can be associated with a process involving $\int dt\int dV$. Note the similarity of system entropies. It indicates that they can be derived in similar ways. Indeed, (\ref{s_thermal}) can be obtained in the same way as obtaining (\ref{S_S_ion}) and (\ref{s_ion}) by thinking $\lambda_z\to a$ and $\varepsilon\to q$ and by dropping the constraint $h=2\lambda_z$. On the other hand, (\ref{s_ion}) can be obtained in the same way as (\ref{s_thermal}).

For the ionic conduction in Fig. \ref{ions}, the competition method can be extended to include, for instance, an extra kind of ions. This means that we can have $\frac{s}{\tau}+\frac{s^{\prime}}{\tau^{\prime}}+\sigma_s = {\it maximum}$ with $\sigma_s=\frac{E}{T}(j+j^{\prime})$. On the other hand, the competition method can be extended to include one more external force, for instance, a concentration gradient. In this case, we have
\begin{equation} \label{sde}
	\frac{s}{\tau}+\sigma_s^{\rm diffusion}+\sigma_s^{\rm conduction}={\rm maximum}.
\end{equation}
Although there are two forces, they act on the same flux, represented by either electric flux or diffusion flux. Here we use diffusion flux. So we have $s=s_0-\frac{1}{2}\frac{k_B\tau^2}{na^2}j^2$ as (\ref{s_diffusion}), $\sigma_s^{\rm diffusion}=-\frac{k_B}{2n}\frac{dn}{dz}j$ as (\ref{sigma_diffusion}), and $\sigma_s^{\rm conduction}=\frac{qE}{T}j$, not (\ref{sigma_s_electric}). Plugging them into (\ref{sde}), we get
\begin{equation} 
	-\frac{1}{2}\frac{k_B\tau}{na^2}j^2-\frac{k_B}{2n}\frac{dn}{dz}j+\frac{qE}{T}j={\rm maximum}
\end{equation}
which leads to
\begin{equation} 
	j=\left(-\frac{dn}{dz}+\frac{2qn}{k_BT}E\right)\frac{a^2}{2\tau}.
\end{equation}
When it is compared with
\begin{equation} 
	j=-D\frac{dn}{dz}+nq\mu E
\end{equation}
we obtain
\begin{equation} \label{Einstein_relation}
	D=\frac{1}{2}k_BT\mu
\end{equation}
where $\mu$ is mobility. The Nernst-Einstein relation is $D=k_BT\mu$. Our result has an extra factor $1/2$. To see how $1/2$ arises, let us do a little statistical analysis: For an ion, during a time interval $\tau$, it can jump either up or down, with the corresponding probability calculated from (\ref{1ion}) as
\begin{equation} 
	P(\uparrow)=\frac{ 	e^{\frac{aqE}{k_BT}} 	}{	e^{\frac{aqE}{k_BT}}+e^{-\frac{aqE}{k_BT}} 	}, \ \ \ \ 
	P(\downarrow)=\frac{ 	e^{-\frac{aqE}{k_BT}} 	}{	e^{\frac{aqE}{k_BT}}+e^{-\frac{aqE}{k_BT}} 	}.
\end{equation}
Let us view the ionic conductor as many layers and focus on only two of them. This means that we move from  Fig. \ref{ions} to Fig. \ref{atoms}. Let the external electric field and the concentration gradient be tuned in balance so that there is no net flux flowing between the two layers: they are in detailed balance,
\begin{equation} 
	n_{\rm up}P(\downarrow)=n_{\rm down}P(\uparrow),
\end{equation}
where $n_{\rm up}$ and $n_{\rm down}$ are the ion concentrations for the up layer and the down layer. Thus we obtain 
\begin{equation}\label{ndownnup} 
	\frac{n_{\rm down}}{n_{\rm up}}=e^{-2\frac{aqE}{k_BT}}.
\end{equation}
The extra factor $2$ means that we are not with the Boltzmann distribution. Yet the Boltzmann distribution is the starting point of deriving the Nernst-Einstein relation. This is where the problem is. Let us continue. We write $\frac{dn}{dz}=\frac{n_{\rm up}-n_{\rm down}}{a}=\frac{n_{\rm up}}{a}(1-\frac{n_{\rm down}}{n_{\rm up}})=\frac{2qn_{\rm up}E}{k_BT}$ where $E\to 0$. Thus the two forces have been tuned in the form $\frac{dn}{dz}=\frac{2qnE}{k_BT}$ where $n$ means $n_{\rm up}$. On the other hand, zero net flux means $-D\frac{dn}{dz}+nq\mu E=0$. Thus, the relation (\ref{Einstein_relation}) is obtained again. The extra factor $1/2$ appears also in work \cite{yjzhang}, as the difference between the known result and the derived ionic conductivity, which can be extracted again from (\ref{j_ion}). But if one uses (\ref{Einstein_relation}) instead of the Nernst-Einstein relation, the known result needs to be re-derived, and will be the same as the other. Nevertheless, we can keep the Nernst-Einstein relation, and call the extra factor 1/2 a Haven ratio 0.5 \cite{Haven1}.

Let us further discuss $P(\uparrow)$ and $P(\downarrow)$. There are two ways to calculate it. Their results are different by the same factor $1/2$. One way is with a classical consideration, the other a quantum consideration. By classical consideration, for an ion to make a jump to a neighboring site, it needs to have an energy that is higher than the potential barrier at position $\frac{1}{2}a$. When an external electric field exists, the potential barrier will change by $\pm\frac{1}{2}aqE$. So the jump probability will change by a factor $\exp\left(\pm\frac{\frac{1}{2}aqE}{k_BT}\right)$. 
 By quantum consideration, when an ion is about to jump, it is first in a superposition state like 
\begin{equation} 
	|\Psi\rangle=	C_{\rm up}|{\uparrow}\rangle+C_{\rm down} |{\downarrow}\rangle
\end{equation}
with $C_{\rm up}$ and $C_{\rm down}$ relating to phase spaces. If the ion is to jump up, it will release energy $aqE$. Then the phase space, which is the number of the final microscopic states, will become greater by a factor $\exp\left(\frac{aqE}{k_BT}\right)$. This means $|C_{\rm up}|^2\propto\exp\left(\frac{aqE}{k_BT}\right)$. In the same way, we have $|C_{\rm down}|^2\propto\exp\left(-\frac{aqE}{k_BT}\right)$. When the ion actually jumps, the superposition state collapses: the ion jumps either up with probability $|C_{\rm up}|^2$ or down with probability $|C_{\rm down}|^2$, as shown in (\ref{1ion}). The results of the two ways are different. But only the result of the quantum consideration leads to the correct entropy production, see (\ref{1ion}), Tab \ref{table} and (\ref{sigma_s_electric}).

\section{conclusion}
We present a competition method, $S_S/\tau +\sigma={\it maximum}$, to calculate the steady flux. The competition method is first formulated by studying ionic conduction based on a statistical analysis. Suppose there are $N$ ions and each ion jumps up and down randomly every time interval $\tau$. The number of ways for $k$ ions to jump up is $\Omega_S(k)=C_N^k$. A $k$ appears with probability $P(k)= \Omega_S(k)/2^N$. The $k$ fluctuates. The electric flux $J_V$ fluctuates too. They relate to each other in the form $J_V= (2k-N)qa/\tau$ where $q$ is the ion charge and $a$ is the lattice constant.

Now we apply an external electric field $E$. The jumps of the ions will be affected. But the effect can be separated out as an entropy production, $\sigma=\frac{E}{T}J_V$. The intrinsic randomness nature of the jumps is left along to be described as a system entropy, $S_S=k_B\ln \Omega_S\propto -J_V^2$, which is now in the position to compete with the entropy production. The competition determines the most probable $J_V$ which maximizes $S_S/\tau +\sigma$. For macroscopic phenomena, the most probable $J_V$ is also the steady $J_V$. We then continue to calculate the flux density $j=J_V/V$ and the ionic conductivity.

The competition method is then applied to other phenomena.

\end{document}